# Partially Identified Treatment Effects for Generalizability

Wendy Chan

*Human Development and Quantitative Methods Division, Graduate School of Education*

*University of Pennsylvania, Philadelphia, Pennsylvania 19104, USA*

(Dated December 12, 2016)

# Partially Identified Treatment Effects for Generalizability

Abstract

Recent methods to improve generalizations from nonrandom samples typically invoke assumptions such as the strong ignorability of sample selection, which is challenging to meet in practice. While researchers acknowledge the difficulty in meeting this assumption, point estimates are still provided and used without considering alternative assumptions. We compare the point identifying assumption of strong ignorability of sample selection with two alternative assumptions—bounded sample variation and monotone treatment response—that partially identify the parameter of interest, yielding interval estimates. Additionally, we explore the role that population data frames play in contributing identifying power for the interval estimates. We situate the comparison around causal generalization with nonrandom samples by applying the assumptions to a cluster randomized trial in education. Bounds on the population average treatment effect are derived under the alternative assumptions and the case when no assumptions are made on the data. While comparing the bounds, we discuss the plausibility of each alternative assumption and the practical tradeoffs. We highlight the importance of thoughtfully considering the role that assumptions play in causal generalization by illustrating the differences in inferences from different assumptions.

Policymakers have become increasingly interested in the extent to which inferences from experimental studies apply to target populations of inference. For example, will the results of a new mathematics curriculum from a sample of schools be applicable when implemented in a larger group of schools? This interest in the external validity or *generalizability* of results reflects the growing use of well-designed evaluations to inform decisions in evidence-based policy making. The generalizability of experimental results to target populations of inference requires both treatment randomization and probability sampling. When probability sampling is used to select the experimental sample, estimates of the treatment effect are both unbiased for the sample and for the population, and generalizations can be made without modelling assumptions. Because the information from experimental studies is typically used to inform policy decisions, particularly in education and medicine, designing evaluations that strengthen both the internal validity and the generalizability of inferences has been particularly important to policymakers and researchers.

In social science research, a common challenge to the generalizability of experimental results is that the samples used in experimental studies are typically not randomly selected; in other words, they do not employ probability sampling (Greenberg & Shroder, 2004; Olsen, Orr, Bell, & Stuart, 2013). The populations of interest are often not specified beforehand or may not be part of the population that is available during the time of the experiment (O'Muircheartaigh & Hedges, 2014). Without probability sampling, generalizing treatment effects is challenging because the bias induced from self-selection no longer allows for model-free estimation of treatment effects (Keiding & Louis, 2016). Statisticians have recently developed methods to improve generalizations from non-probability samples using propensity scores (Stuart, Cole, Bradshaw & Leaf, 2011; Tipton, 2013; O'Muircheartaigh & Hedges, 2014). Propensity score methods match experimental samples to an inference population based on observable characteristics so that the matched groups are compositionally similar (Rosenbaum & Rubin, 1984). This application extends previous work on the use of propensity scores in quasi-experimental and observational studies that addressed treatment selection bias (Rosenbaum & Rubin, 1983). These methods also extend those used to adjust for non-response in survey sampling (Little, 1986). Post-hoc adjustments using propensity scores include inverse propensity weighting (IPW; Stuart et al., 2011) and subclassification (Tipton, 2013; O'Muircheartaigh & Hedges, 2014), both of which reweight the sample to obtain bias-reduced estimates of the average treatment effect.

While propensity score methods have significantly contributed to causal generalization, the assumptions required for these methods are often difficult to meet and controversial. In particular, the generalizability of treatment effects requires that sample selection be *strongly ignorable* given the propensity scores when the sample is not randomly selected. Strong ignorability of sample selection requires that two conditions be met. First, the propensity scores must contain information on all possible covariates that explain treatment effect variation and affect sample selection. If this is met, when propensity scores are included in the analysis, the sample can be considered like a probability sample. Second, every unit in the population must have a "similar unit" in the experimental sample in which the comparability is based on the covariates used in the propensity score model. Situations in which strong ignorability of sample selection fails for either one or both conditions can occur, and it is important for researchers to consider the applicability of results when core assumptions do not hold.

In practice, assumptions like strong ignorability of sample selection are made in order to get *point estimates* (single values) of average treatment effects. However, an important concern is that while researchers acknowledge the challenges in meeting core assumptions and empirically validating them, point estimates are still given and used in spite of the challenges (Stuart et al., 2011; Tipton, Hedges, Vaden-Kiernan, Borman, Sullivan, & Caverly, 2014). Importantly, alternative assumptions to strong ignorability of sample selection may provide different perspectives on the same problem. Since different assumptions can lead to different inferences, it is essential for researchers to consider the plausibility of different assumptions and the role that assumptions play in the resulting inferences. For generalization in particular, if strong ignorability of sample selection is not met, researchers should discuss alternative assumptions and their tradeoffs.

The goal of this article is two-fold. First, we illustrate that some assumptions, not necessarily strong ignorability of sample selection, are *necessary* for generalizations to be informative in determining the direction of the expected treatment effect. This is particularly true for generalizations from small experimental studies in which the results are applied to a population that is at least 20 times larger than the sample. We demonstrate that this is also the case even if auxiliary sources of data on the population (as is commonly used for generalizations) are available. Second, we explore and compare alternative assumptions to strong ignorability of sample selection and consider their practical trade-offs. In particular, we focus on bounded sample variation and monotone treatment response, which differ from strong ignorability of sample selection as they are insufficient to point identify the average treatment effect and instead lead to *interval estimates* of the population parameter. The fact that these assumptions lead to a range of possible values and impose few restrictions on the distribution of outcomes has led researchers to refer to them as weaker assumptions (Manski, 2009).

We center our discussion of these alternative assumptions around a completed cluster randomized trial (CRT) in education. We focus on this CRT because the process by which schools were recruited is common among educational CRTs, the features of the study (sample to population size ratio) are typical of many educational studies, and initial analyses of the sample and population data suggest that strong ignorability of sample selection may not hold. The preliminary analysis of the results was completed in Konstantopoulous, Miller, and van der Ploeg (2013) and the generalizability of point estimates of the treatment effect was considered in Tipton, Hallberg, Hedges, and Chan (2016). In order to understand the roles and tradeoffs of different assumptions, we pose three important questions. First, are interval estimates of average treatment effects informative when no assumptions are made on the experimental data? This question is addressed by examining the interval estimates from the CRT solely based on the data and the randomized nature of the treatment assignment. Second, can auxiliary sources of data improve upon the interval estimates of treatment effects in causal generalization? This question is addressed by considering differences in the interval estimates of treatment effects when covariate information from the population is used to contribute identifying power. Third, what inferences can be made when bounded sample variation and monotone treatment response are imposed, and what are the tradeoffs among these assumptions? To investigate this issue, we present the bounds for the CRT under the two alternative assumptions, discuss the plausibility of each assumption for the CRT, and compare the interval estimates to the ones under the no-assumptions framework. We also compare each set of interval estimates with the point estimates derived under strong ignorability of sample selection.

Although standard errors are provided for the point estimates, they are not provided for the interval estimates for two reasons. First, because our discussion centers on identification of treatment effects when strong ignorability of sample selection is not met, we focus our comparison between the nonparametric interval estimates and the point estimates. Second, measures of statistical precision require specification of the sampling process that generated the data, which is unknown in the case of generalization studies from non-probability samples (Manski & Pepper, 2015). We instead compare the inferences from different assumptions by assessing the differences in magnitudes of the estimates.

The article is organized as follows. In the first section, we present the CRT example and discuss the generalization problem of interest. In the second section, we introduce the notation and assumptions needed for point identification of average treatment effects and discuss the plausibility of strong ignorability of sample selection. In the third section, we introduce the frameworks for deriving bounds and provide the bounds for the no-assumptions case (under treatment randomization), bounded sample variation, and monotone treatment response. Additionally, a "fusion" approach combining the bounding methods with propensity score subclassification is explored. In the final two sections, we apply the bounds to the CRT, discuss the plausibility of each alternative assumption and the tradeoffs, and compare these bounds with the point estimates of the average treatment effect. We conclude with a discussion of the role of assumptions in causal generalization and ideas for future research.

**CRT Example**

In 2006, the Indiana Department of Education and the Indiana State Board of Education managed the implementation of a new assessment system to measure annual student growth and to provide feedback to teachers (Konstantopoulos et al., 2013). During the 2009-2010 academic year, 56 K–8 (elementary to middle) schools from the state of Indiana volunteered to implement the new system, of which 34 were randomly assigned to the state's new assessment system while 22 served as control schools. In the treatment schools, students were given four diagnostic assessments that were aligned with the Indiana state test, and their teachers received online reports on their performance to dynamically guide their instruction in the periods leading up to the state exam. The effectiveness of the assessment system was measured using the Indiana Statewide Testing for Educational Progress-Plus (ISTEP+) scores in English Language Arts (ELA) and mathematics. For each study school, the ISTEP+ scores were discretized using the minimum cutoff scores from the Indiana Department of Education and aggregated as either "Pass" or "Not Pass."

    A natural question emerges from this study: If every school in Indiana were to implement this system, what is the expected impact on student achievement? In other words, to what extent do the results from the Indiana CRT generalize to the entire state? If the study was planned with generalization in mind, both treatment randomization and probability sampling would be implemented to facilitate causal generalization by directly estimating the average treatment effect for the population of Indiana schools. However, a key concern for generalization in this example is whether this sample of volunteer schools is "representative" of the 1,514 K–8 schools in Indiana during that year (Kruskal & Mosteller, 1980). If the schools that self-selected into this study differed from the schools that did not volunteer, (for example, on the demographic composition of its students or in the schools' past performance), and these differences moderated potential treatment effects, any estimate of the treatment effect of the benchmark assessment system will be biased for the population of Indiana schools. By making assumptions about the expected treatment effects among the volunteer schools and the schools not in the experimental study, it is possible to generate bias-reduced estimates of the average treatment effect for the

population. In the next section, we introduce notation and formally define the assumptions needed to get point estimates of the average treatment effect.

**Notation and Assumptions**

The estimation of causal treatment effects is framed using Rubin's Causal Model (Rubin 1974, 1977, 1980, 1986). Let *P* denote the population of inference consisting of *N* schools, of which *n* schools are selected into the sample. Let *W* be an indicator of treatment assignment where *W = 1* if a school was assigned to implement the assessment system (treatment) and *W = 0* if a school was not assigned to implement the system (control). For each school in *P*, let *Y(W)* denote the binary potential outcome of whether a school received a "Pass" score or not under the respective treatment condition (*W=0,1*). Finally, let *Z* be an indicator of sample selection where *Z = 1* if a school was in the experimental study and *Z = 0* otherwise.

To estimate treatment effects, the stable unit treatment value assumption (SUTVA) for the sample and population is required (Rubin, 1978, 1980, 1990; Tipton, 2013). Under SUTVA for the sample, there is only one version of the treatment, and the potential outcomes *Y(0), Y(1)* of each school depend only on the treatment received by that school and not on the treatment received by another school. Under SUTVA for the population, this must also hold for the sample selection process, in which the potential outcomes do not depend on the proportion of schools selected into the experiment and the potential treatment effects do not depend on being involved in the experimental study. Additionally, SUTVA requires that there is no interference between schools, both between the treatment and control schools in the sample, and between the volunteer and non-volunteer schools in the population.

Assuming that SUTVA holds, the treatment effect for each school in the sample and population (the volunteer and non-volunteer schools) is defined as τ = *Y(1) - Y(0)*, and because *P* consists of schools that were selected and not selected into the experimental study, we can define two average treatment effects

(Sample): $\tau_{SATE} = E(\tau|Z=1)$

(Population): $\tau_{PATE} = E(\tau|Z=1) * P(Z=1) + E(\tau|Z=0) * (1-P(Z=1))$

where $\tau_{SATE}$ is the expected sample average treatment effect (SATE) and $\tau_{PATE}$ is the expected population average treatment effect (PATE). For generalization, the PATE is the parameter of interest. Note that the PATE defined here is the measure of impact for all schools in the state of Indiana, but a different PATE can be estimated for populations or groups that change over time and space. Importantly, the SATE and PATE are equivalent when $E(\tau|Z=1) = E(\tau|Z=0)$, the case under probability sampling, when the potential treatment effects are constant, when every school in the population is in the study ($P(Z=1)=1$), or when sample selection and heterogeneity in treatment effects are independent (Rubin, 1974; Imai, King, & Stuart, 2008). Otherwise, estimates of the SATE are considered to be naïve and biased estimates of the PATE.

For each school in $P$, we assume that a vector of characteristics or covariates, **X**, is observed where **X** may include both categorical and continuous measures. This covariate information is used to compare the volunteer with the non-volunteer schools in the study and can be obtained from sources such as the Common Core of Data or state longitudinal data systems. For each school, the sampling propensity score (for a finite population) is defined as:

$$s(\mathbf{X}) = P(Z=1|\mathbf{X})$$

Propensity scores model the probability of sample membership as a function of **X** and have the advantage of being balancing scores where matching by the propensity score is equivalent to matching by the covariates in the propensity score model (Rosenbaum & Rubin, 1983). A common method of estimating $s(\mathbf{X})$ is with a logistic regression model with an intercept term $\beta_0$,

$$\log(s(\mathbf{X})/(1-s(\mathbf{X}))) = \beta_0 + \beta_1 X_1 + \beta_2 X_2 + \cdots + \beta_p X_p$$

based on $\mathbf{X}=(X_1, X_2, \ldots, X_p)$ covariates.

To obtain bias-reduced point estimates of the PATE using propensity scores, several assumptions, in addition to SUTVA, are needed.

First, treatment assignment must be strongly ignorable given the propensity scores (Stuart et al., 2011; Tipton, 2013):

$$Y(1), Y(0) \perp W \mid Z=1, s(\mathbf{X}) \text{ and } 0 < P(W=1|Z=1, s(\mathbf{X})) < 1$$

Among schools selected into the study ($Z=1$), the potential outcomes are conditionally independent of treatment assignment, and every school in the study must have some probability of being assigned to the treatment condition. This condition is typically met in randomized experiments such as the Indiana CRT (Rosenbaum & Rubin, 1983; Stuart et al., 2011).

Second, unconfounded sample selection must hold where sample selection is conditionally independent of the treatment effects (Stuart et al., 2011; Tipton, 2013):

$$(\tau = Y(1) - Y(0)) \perp Z \mid s(\mathbf{X})$$

Finally, strong ignorability of sample selection is needed to fully identify the PATE. Because this assumption is the focus of this article, we refer to this assumption as simply sampling ignorability hereafter. This assumption requires both unconfounded sample selection and that the distribution of covariates $\mathbf{X}$ in the sample and population share common support (Tipton 2013, 2014):

$$\tau \perp Z \mid s(\mathbf{X}) \text{ and } 0 < s(\mathbf{X}) \leq 1$$

The first stipulation, unconfounded sample selection, requires that $\mathbf{X}$ (and consequently $s(\mathbf{X})$) includes all covariates that explain the potential variation in treatment effects and sample selection. The second stipulation requires that every school in the population must have a relevant comparison school in the sample so that no school should have s(**X**)=0.

**Sampling Ignorability in the Indiana CRT**

Whether the sampling ignorability assumption is credible and plausible in practice is a controversial topic. At the heart of the matter, sampling ignorability is an invariance assumption in which the effect of the assessment system is the same (invariant) for students, on average, regardless of whether the school volunteered to be in the Indiana CRT, once the propensity scores are taken into account. In other words, if sampling ignorability holds, self-selection does not matter because any differences between the volunteer and non-volunteer schools are explained by the propensity scores. Conceivably, sampling ignorability may not hold for a few reasons. For example, schools that respond differently to the assessment system may have a strong support base from parents, which may not be measured and would therefore be omitted from **X**. Additionally, some schools that chose not to volunteer in the Indiana CRT may have an assessment system already in place and the impact of the CRT system may be different for these schools compared with schools that did not have such a system or such resources to begin with for their students. If these characteristics of schools are not included in **X**, sampling ignorability does not hold. Note that the concern here is that these potential covariates may explain treatment effect variability, but they are not included in the propensity score model. Alternatively, if the Indiana sample consisted of all single-gender schools, but generalization was made to a population of co-educational schools, sampling ignorability does not hold. Co-educational schools in the population would not have appropriate matches in the sample so that these schools would have a sampling propensity score s(**X**) ≈ 0.

**Partial Identification of the PATE**

Examples in which sampling ignorability fails, as described above, can occur in practice. Manski (2009) first recommended that researchers begin analyses by considering what can be learned from the data alone, without any assumptions, so that a "domain of consensus" is the established starting point. If sampling ignorability is unlikely to hold, other assumptions that *partially* identify the PATE, yielding interval estimates, can provide alternatives. Partial identification methods were first developed in response to the concern that the credibility of assumptions was compromised by the use of strong point-identifying assumptions, many of which were clearly violated in practice (Manski, 1990). Instead of using point-identifying assumptions, partial identification analyzes the extent to which alternative assumptions yield potentially informative bounds.

To explore these partially identifying assumptions, we first decompose the estimator of the PATE and consider the role that assumptions play in contributing identifying power. Because our empirical example is a CRT, we frame this analysis around a randomized experiment and assume SUTVA, strongly ignorable treatment assignment, and perfect compliance. The average treatment effect, $E(\tau) = E(Y(1) - Y(0)) = E(Y(1)) - E(Y(0))$, is the difference of two expected potential outcomes. Using the law of iterated expectations, the SATE, a function of the treatment assignment indicator $W$, is decomposed as follows:

$$E(Y(1)) = E(Y(1)/W=1) * P(W=1) + E(Y(1)/W=0) * P(W=0) \qquad (1)$$
$$E(Y(0)) = E(Y(0)/W=1) * P(W=1) + E(Y(0)/W=0) * P(W=0) \qquad (2)$$

The PATE, a function of both $W$ and sample selection indicator $Z$, is decomposed as follows:

$$E(Y(1)) = E(Y(1)/W=1, Z=1) * P(W=1, Z=1) + E(Y(1)/W=0, Z=1) * P(W=0, Z=1) +$$
$$E(Y(1)/W=1, Z=0) * P(W=1, Z=0) + E(Y(1)/W=0, Z=0) * P(W=0, Z=0) \qquad (3)$$

$$E(Y(0)) = E(Y(0)/W=1, Z=1) * P(W=1, Z=1) + E(Y(0)/W=0, Z=1) * P(W=0, Z=1) +$$
$$E(Y(0)/W=1, Z=0) * P(W=1, Z=0) + E(Y(0)/W=0, Z=0) * P(W=0, Z=0) \qquad (4)$$

Since each study school is assigned to at most one treatment, the quantities in Equations (1) through (4) cannot be identified, a premise of the Fundamental Problem of Causal Inference (Holland, 1986). The terms E(*Y(1)*/*W=0*) and E(*Y(0)*/*W=1*) are unobservable counterfactuals since they refer to the expected outcome under treatment (control) when assigned to control (treatment), which are unknown (Greenland, Pearl, & Robins, 1999; Dawid, 2000). We refer to these quantities as treatment counterfactuals.

The PATE is a function of the same treatment counterfactuals so it, too, cannot be identified. However, the decomposition of the PATE differs from that of the SATE in two important ways. First, the potential outcomes in (3) and (4) necessarily include additional counterfactual terms because the PATE requires information about sample selection, given by *Z*. Second, the PATE includes four additional counterfactuals, E(*Y(1)*/*W=1, Z=0*), E(*Y(1)*/*W=0, Z=0*), E(*Y(0)*/*W=1, Z=0*), and E(*Y(0)*/*W=0, Z=0*), which will be referred to as the *sample counterfactuals*. These are the potential outcomes under a treatment condition for schools not selected into the experimental study. Note that the treatment and sample counterfactuals are unobservable for different reasons. The goal of any causal inference study is to identify both types of counterfactuals, whether through the design stage of the study or through the use of assumptions on the distribution of these potential outcomes (Rubin, 2011).

Since the Indiana CRT was a randomized experiment, treatment assignment is strongly ignorable so that *Y(1), Y(0)* ⊥ *W*, E(*Y*/*W=1*)= E(*Y(1)*/*W=1*) = E(*Y(1)*)) and E(*Y*/*W=0*)= E(*Y(0)*/*W=0*) = E(*Y(0)*)) and the SATE is point identified (Rubin, 1978). Under treatment randomization, the distribution of unobservable treatment counterfactuals is equivalent to that of the realized potential outcomes, which allows for model-free estimation of the SATE (Rubin, 1974; Imai, King, & Stuart, 2008). For the PATE, however, the four sample counterfactuals remain unobservable in the absence of probability sampling, so that assumptions on the distribution of these counterfactuals are needed in order to achieve point identification.

In the following sections, we exclude sampling ignorability and begin with estimates of the PATE based on the data and known features of the experimental outcome. Because the ISTEP+ scores in the Indiana example were aggregated into "Pass" and "Not Pass," the subsequent bounds are derived using a binary *Y*. The potential outcomes *Y(0), Y(1),* therefore share the same lower and upper bound, {*0,1*}, for all units in *P* and the expectations E(Y(1)), E(Y(0)) become the probabilities P(*Y(1)=1*), P(*Y(0)=1*). These bounds can easily be extended to bounded continuous *Y*, such as test scores, where the lower and upper bounds of *Y* are used in place of {0,1}. Cases in which *Y* is bounded on one side, but unbounded on the other have been discussed in Manski (2009), though their focus is on experimental studies and not causal generalization.

**Two Frameworks for Estimating Bounds on the PATE**

For generalizability, a population data frame is required in order to estimate the propensity score of being selected into the experimental sample for every school in *P* (Tipton et al., 2014). The population data frame used in the Indiana CRT, sourced from the Common Core of Data and the Indiana Department of Education, contains demographic information on students and schools as well as test scores over several years. Since these data frames enumerate all schools in *P*, they provide information on the sample counterfactuals in the decomposition of (3) and (4). While propensity score methods use the population data to model the selection probability, we propose using the data frame to present two frameworks for estimating bounds of the PATE.

We introduce the *full-interval* framework and the *reduced-interval* framework, which differ by the extent to which the population data frame is useful for providing information to tighten the bounds. The full-interval framework is solely based on the experimental sample data and makes no assumptions on the sample counterfactuals. Under this framework, the only observable quantities are the realized potential outcomes E(*Y(1)*/*W=1, Z=1*) and E(*Y(0)*/*W=0, Z=1*), while the unobservable treatment and sample counterfactuals are replaced by known bounds on the outcome. The reduced-interval framework uses the empirical evidence from the experimental sample and the population data frame (that is, both the study data and population data from the Common Core of Data) to identify the sample counterfactual E(*Y(0)*/*W=0, Z=0*). This counterfactual represents the expected outcome under the control condition for schools that were assigned to the control group *(W=0)* and that were not in the experimental sample *(Z=0)*.

A rationale for the use of the reduced-interval framework lies in the idea that the control condition in educational experiments may be a "business as usual" condition, where control schools continue implementing existing curricula or programs. The reduced-interval framework considers the distribution of potential outcomes among schools not selected into the experiment (that is, *Z=0*) to be identified by the population data frame if the control condition was "business as usual." Because this was the case for the Indiana CRT, we argue that the non-sample schools in the population were similarly exposed to the control condition so that their potential outcomes under control are identified by the population data frame. We compare the widths and magnitudes of the estimated bounds of the PATE under these two frameworks to assess the identifying power of the population data frame on the interval estimates. Note that the reduced interval framework requires additional information, specifically on the probabilities P(*W=1*/*Z=0*), P(*W=0*/*Z=0*), and P(*Z=0*). These three probabilities are all functions of P(*Z=0*), which is estimated by the proportion of the population not selected into the experimental sample.

**Bounds Under Treatment Randomization**

Using the full- and reduced-interval frameworks, we begin by estimating the "worst case" bounds for binary outcomes under treatment randomization using the data alone (Manski, 2009). From (3) and (4), the lower and upper bounds for the potential outcomes are derived by replacing the unobservable sample counterfactuals with 0 and 1, respectively. No substitution is made for the treatment counterfactuals because the potential outcomes are statistically independent of the treatment indicator *W*.

**Full-Interval Framework**

Under this framework, the bounds associated with treatment *w*, where *w* ∈ *W*, are:

$$E(Y(w)) \in [E^L(Y(w)), E^U(Y(w))] \qquad (5)$$

$$E^L(Y(w)) = E(Y(w)/W=w, Z=1) * P(Z=1)$$

$$E^U(Y(w)) = E^L(Y(w)) + (1-P(Z=1))$$

**Reduced-Interval Framework**

When the population data frame identifies the sample counterfactual E(*Y(0)*/*W=0, Z=0*), no replacements are made for this potential outcome. Because this sample counterfactual is the expected outcome under control, the lower and upper bounds for E(*Y(1)*) remain the same, but the bounds for E(*Y(0)*) become

$$E(Y(0)) \in [E^L(Y(0)), E^U(Y(0))] \qquad (6)$$

$$E^L(Y(0)) = E(Y(0)/W=0, Z=1) * P(Z=1) + E(Y(0)/W=0, Z=0)*P(W=0, Z=0)$$

$$E^U(Y(0)) = E^L(Y(0)) + (1-P(Z=1) - P(W=0, Z=0))$$

**Bounds on the PATE**

For both frameworks, the lower and upper bound of the PATE are given by the differences

$$PATE^L = E^L(Y(1))-E^U(Y(0)) \qquad (7)$$

$$PATE^U = E^U(Y(1))-E^L(Y(0))$$

Under the given information structure, the bounds in (7) under treatment randomization are sharp (see Online Supplementary Materials for proof). The width of the bound in (5) is 2* P(*Z=0*), which is strictly smaller than 1 when P(*Z=0*) < 0.5. The width of the bound using (6) changes to P(*Z=0*) + P(*W=1, Z=0*) under the reduced-interval framework, which is smaller than and at most equal to the width for the full-interval case. Because of the smaller width, the population data frame has the potential to contribute identifying power and tighten bounds.

While the bounds in (5) and (6) provide the simplest (nonparametric) interval estimates of the PATE, they are rarely informative for identifying the sign of the treatment effect. In studies like the Indiana CRT, the sample is generalized to a population at least 20 times larger, so that the probability of not being in the sample, P(*Z=0*), is likely to be greater than 0.5, resulting in bounds that include zero (signifying an insignificant PATE). Assumptions are thus necessary for generalizations from studies that exhibit similar sample size ratios. In the next sections, we introduce the bounded sample variation and monotone treatment response assumptions and compare the widths of the resulting bounds to that of (5) and (6) when no assumptions are made.

**Bounded Sample Variation and Treatment Randomization**

Bounded variation assumptions were first introduced in Manski (2015) and have been discussed in Manski and Pepper (2015) with applications to the impact of right-to-carry laws on crime rates. Unlike sampling ignorability, bounded sample variation is an assumption made on the expected *sample counterfactuals*, not treatment effects. In particular, this assumption stipulates that the outcomes E(*Y(1)/W=1, Z=0*), E(*Y(1)/W=0, Z=0*), E(*Y(0)/W=1, Z=0*), and E(*Y(0)/W=0, Z=0*) are "similar" to the observable, realized outcomes among the sample schools. We quantify this similarity for randomized experiments as follows:

$$|E(Y(w)/W=w, Z=0) - E(Y(w)/W=w, Z=1)| \leq \lambda \qquad (8)$$
$$|E(Y(w)|W \neq w, Z=0) - E(Y(w)|W \neq w, Z=1)| \leq \lambda \text{ for } w \in W$$

Here, $\lambda \in [0,1]$ is a constant that represents the largest magnitude of the absolute difference (Manski, 2015). Note that the condition in (8) is applied to both sets of sample counterfactuals, and no substitution is made for the treatment counterfactuals under treatment randomization. By design, bounded sample variation yields interval estimates of the PATE where the width of the intervals is based on $\lambda$. For the Indiana CRT, if bounded sample variation holds, the proportion of "Pass" schools differs by at most a constant $\lambda$ between the volunteer and non-volunteer schools for each respective treatment condition.

Under this assumption, for $w \in W$, the bounds of the PATE are given by

$E^L(Y(w)) = E(Y(w)/W=w, Z=1) * P(Z=1) + (E(Y(w)/W=w, Z=1) - \lambda) * P(Z=0)$ (9)

$E^U(Y(w)) = E(Y(w)/W=w, Z=1) * P(Z=1) + (E(Y(w)|W=w, Z=1) + \lambda) * P(Z=0)$

for the full-interval framework and

$E^L(Y(0)) = E(Y(0)/W=0, Z=1) * P(Z=1) + E(Y(0)/W=0, Z=0) * p + (E(Y(0)/W=0, Z=1) - \lambda) * (1 - P(Z=1) - p)$ (10)

$E^U(Y(0)) = E(Y(0)/W=0, Z=1) * P(Z=1) + E(Y(0)/W=0, Z=0) * p + (E(Y(0)/W=0, Z=1) + \lambda) * (1 - P(Z=1) - p)$

for the reduced-interval framework, where $p = P(W=0, Z=0)$. Like the previous frameworks, the lower and upper bounds of the PATE are given by (7). Because $Y$ is binary, the constant $\lambda$ lies in the interval $[0,1]$, as it represents a difference in estimated proportions. However, $\lambda$ can be any positive constant when $Y$ is continuous. Larger values of $\lambda$ imply larger differences between the expected potential outcomes, thus widening the interval estimates and weakening the assumption further by allowing more flexibility in the absolute differences in potential outcomes. The bounds in (9) improve upon the worst case bounds in (7) for binary $Y$ and are sharp (see Online Supplementary Materials) if

$E(Y(1)/W=1, Z=1) - E(Y(0)/W=0, Z=1) + 2\lambda < 1$ and
$E(Y(1)/W=1, Z=1) - E(Y(0)/W=0, Z=1) - 2\lambda > -1$

Since each sample counterfactual $E(1)/W=1, Z=0)$, $E(Y(1)/W=0, Z=0)$, $E(Y(0)/W=1, Z=0)$, and $E(Y(0)/W=0, Z=0)$ is replaced by an expression based on $\lambda$ rather than zero (one) for the lower (upper) bound, the bounds under bounded sample variation are narrower for certain ranges of $\lambda$.

Importantly, when $\lambda = 0$ is chosen for both treatment conditions, the bounds reduce to a single value of the PATE, and point estimation is recovered. Choosing $\lambda$ to be zero implies that the expected proportion of "Pass" schools under treatment (or control) is the same, regardless of whether the school was selected into the study. This choice of $\lambda$ also implies that the distributions of the expected outcome among schools in the sample and schools not in the sample are equivalent for each treatment condition. Note that bounded sample variation and sampling ignorability are not nested. Bounded sample variation allows the potential outcomes among sample and non-sample schools to vary under the assumption that the outcomes are similar in magnitude but not necessarily equal.

It is helpful to conceptualize the plausibility of this assumption in the context of matching. When an experimental sample is matched to a population $P$ in causal generalization, the goal is to achieve balance among observable covariates so that the resulting differences in distributions is minimized (see, e.g., Hansen [2004] for examples of matching). Here, balance is quantified as attaining the smallest standardized mean difference among covariates between the two groups. Conceivably, the difference in expected potential outcomes is smaller with matched samples if the potential outcomes $Y(1), Y(0),$ are a function of the covariates that are balanced between the groups. The plausibility of bounded sample variation lies in the assumption that there is sufficient overlap between the distributions of potential outcomes among sample and non-sample schools to facilitate the derivation of informative bounds.

*Choice of* $\lambda$

Bounded sample variation assumptions require the researcher to choose $\lambda$. In this section, we provide several data-based suggestions for this parameter. Previous work on bounded variation assumptions estimate $\lambda$ based on prior outcome data (Manski & Pepper, 2015).

However, estimates of $\lambda$ from prior data are difficult in our example as it represents the difference between the volunteer and non-volunteer schools' outcomes in the Indiana CRT, which is specific to this study. Because bounded sample variation is related to the goals of matching methods, one natural choice is the absolute standardized mean difference (ASMD) of observable covariates. In particular, let $\lambda = |\mu_P - \bar{X}_S|/\sigma_P$, where $X$ is a pre-treatment covariate, $\mu_P$ is the mean covariate value for schools in the population, $\bar{X}_S$ is the mean covariate value for the sample schools, and $\sigma_P$ is the standard deviation of $X$ across all schools in the population. This choice of parameter reflects the assumption that $\lambda$ effectively measures the degree to which the sample and population is balanced on a specific covariate. As a result, larger values of the ASMD suggest larger differences between the sample and population distributions of the covariate, which is then reflected in wider intervals. To give an example, suppose $Y$ is a test score outcome. One possible choice for $X$ is a pre-test score that is strongly correlated with $Y$. The interval estimate of the population parameter is then determined by the extent to which the sample and non-sample schools are balanced on the pre-test scores $X$. Alternatively, because the ASMD is typically estimated using multiple covariates, other choices of $\lambda$ are the average ASMD of several covariates or the maximum of the ASMD to use as a conservative estimate. Using the ASMD for $\lambda$ allows researchers to base the parameter on empirically derived balance statistics.

A caveat of using the ASMD for $\lambda$ is that values greater than one are possible. In these cases, the bounds under bounded sample variation and treatment randomization may not improve upon the worst case bounds so that the direction of the PATE is again unidentified. However, it is important to compare the bounds from different choices of $\lambda$ when the bounds are tighter than those under the no-assumptions case. Comparing the bounds from different values of $\lambda$ to the case when $\lambda = 0$ allows researchers to assess how balance between the sample and population changes the inferences from the interval estimates.

Aside from the ASMD, another possible choice for $\lambda$ is based on the actual variation of the realized potential outcomes in the experimental study. In particular, let $\lambda = 2*\sqrt{(\text{Var}(Y|Z=1))}$ where $\lambda$ is analogous to the margin of error in the construction of confidence intervals for normally distributed data. When $Y$ is binary, the variance is a function of the proportion of "successes" in the sample (that is, the proportion of "Pass" schools). Choosing $\lambda$ in this way offers a conservative limit on the difference in $E(Y)$ by setting the difference to be no more than two standard deviations from the empirical distribution of outcomes. Note that this choice of $\lambda$ is based on the realized potential outcomes in the sample. If outcome data is available for all schools in the population, $\lambda$ can be estimated using information from all of the schools. Alternatively, if $\lambda$ is based on the sample alone and the variance estimates differ among the treatment groups, a conservative choice of $\lambda$ is the maximum of the variance estimates.

Importantly, bounded sample variation is not necessarily validated using these choices of $\lambda$. We provide these suggestions as a starting point for applications of this assumption if the assumption is plausible in a given study. The suggestions for $\lambda$ here are not exhaustive and additional choices based on the balance between propensity score logits (logit(s(**X**)) = log(s(**X**)/(1-s(**X**)))) offer other options when invoking bounded sample variation assumptions.

**Monotone Treatment Response**

In some cases, the researcher may have prior knowledge on an intervention and be confident in its positive effect on outcomes of interest. If the principal investigators of the Indiana CRT were confident in and believed that the benchmark assessment system improved student outcomes, this would lead to a different assumption with a different set of bounds. Under *monotone treatment response* (MTR), the response variable *Y* is said to vary monotonically if, given two treatments *w, w'* ∈ *W*, the following condition holds:

$$w \geq w' \rightarrow Y(w) \geq Y(w')$$

We assume that this condition holds for all schools in the population (that is, those in the sample (*Z=1*) and not in the sample (*Z=0*)). For the Indiana CRT, let *w* = 1 and *w'* = 0 so that under MTR, the proportion of "Pass" schools in the treatment group is assumed to be at least large as the proportion of "Pass" schools in the control group. MTR is related to the idea that the benchmark assessment system at least "does no harm" and at worst, that the expected outcome under treatment is not significantly different from the expected outcome under "business as usual."

Like bounded sample variation, MTR differs from sampling ignorability because it is not an assumption on the treatment effect. Instead, MTR is an assumption made on the response function *Y*. MTR differs from both bounded sample variation and sampling ignorability on two important aspects. First, the treatment indicator *W* now denotes an *ordered* set of treatments. Second, the bounds under MTR are derived using *realized* values of *Y* across all schools in the study. The original MTR framework proposed in Manski (1997) assumed that outcomes at different levels of the treatment were observable, which contributed identifying power even if some levels of the treatment were not realized. A common application of this framework is in labor economics, where each individual's wage function in the labor market is a monotonic function of years of schooling (Manski, 2009). The sharpness of the bounds under MTR is a consequence of the monotonic nature of the response function. Although the focus here is on weakly increasing response functions *Y*, the results can easily be generalized to weakly decreasing functions.

If MTR holds, this assumption, combined with SUTVA, strongly ignorable treatment assignment, and perfect compliance, yields a new set of bounds of the PATE with analogous extensions to the bounds of the prior frameworks. If *Y* is weakly increasing in *W,* the lower bound of the PATE is zero by design. For binary *Y* and *W,* the upper bound is a function of the proportion of "Pass" and "Not Pass" schools in the sample and is given by

$$\text{PATE}^U = P(Y=0|W=0, Z=1) * P(W=0, Z=1) + P(Y=1|W=1, Z=1) * P(W=1, Z=1) +$$
$$P(Y=0|W=0, Z=0) * P(W=0, Z=0) + P(Y=1|W=1, Z=0) * P(W=1, Z=0) \qquad (11)$$

The bounds [0, $\text{PATE}^U$] are tight under the MTR assumption where *Y* is a weakly increasing function of the treatment *W* for all schools in the population (see Online Supplementary Materials). The upper bound is the sum of two components. If *Y* is monotone in treatment, the largest feasible upper bound is the sum of the upper bound for *Y(1)* and the lower bound for *Y(0)*. This is given in the first term, $P(Y=0|W=0, Z=1) * P(W=0, Z=1) + P(Y=1|W=1, Z=1) * P(W=1, Z=1)$, which is the sum of the proportion of "Not Pass" schools (*Y=0*) among the sample schools assigned to control (*W=0*) and the proportion of "Pass" schools (*Y=1*) among the sample schools assigned to treatment (*W=1*). The second term, $P(Y=0|W=0, Z=0) * P(W=0, Z=0) + P(Y=1|W=1, Z=0) * P(W=1, Z=0)$, is the sum of the analogous proportions among the non-sample schools (*Z=0*).

Because MTR is based on realized values *Y*, additional consideration must be taken for the upper bound since it is a function of schools not selected into the experimental sample. If the population data frame contributes identifying power, as proposed under the previous reduced-interval framework, the term *P(Y=0|W=0, Z=0)* is identified using the empirical evidence. However, the proportion *P(Y=1|W=1, Z=0)* is still an unobservable sample counterfactual, and the only information that contributes identifying power are the known bounds {*0,1*} for the outcome *Y*. Because the upper bound depends on this sample counterfactual, an interval of values for PATE$^U$ can be derived by substituting 0 (1) for the minimum (maximum) upper PATE bound. Although we provide both bounds for PATE$^U$ in the results section, we focus on the minimum of the PATE$^U$ as it is derived based on the data alone using the realized values *Y*. This lower bound of the PATE$^U$ is a "best case" bound because it represents the smallest feasible value of PATE$^U$ under the MTR framework. Thus, the smallest PATE$^U$ is derived by substituting 0 for the unobserved sample counterfactuals and the largest PATE$^U$ is derived by substituting the value 1. Like the full- and reduced-interval cases, two MTR bounds, denoted as the "sample MTR bound" and the "population MTR bound," can be written:

*Sample MTR Bound*

PATE$^U$ = *P(Y=0/ W=0, Z=1)* * *P(W=0, Z=1)* + *P(Y= 1/ W=1, Z=1)* * *P(W=1, Z=1)*

*Population MTR Bound*

PATE$^U$ = *P(Y=0/ W=0, Z=1)* * *P(W=0, Z=1)* + *P(Y= 1/ W=1, Z=1)* * *P(W=1, Z=1)* +
    *P(Y=0/ W=0, Z=0)* * *P(W=0, Z=0)*                                                (12)

The bounds in (12) differ by the probability *P(Y=0/W=0, Z=0)*, which is identified by the population data frame in the "population MTR bound." As a result, if MTR holds and the treatment is believed to at least "do no harm," the interval estimate of the PATE shrinks to lie to one side of zero.

**Bounds by Propensity Score Stratum**

The bounds estimated thus far are based on the entire experimental sample and population data frame (the latter for the reduced-interval and population MTR cases). In generalizations with nonrandom samples, a primary goal in the study is to match the sample and population. Subclassification is a common matching method in which the population is partitioned into smaller subclasses or strata using quintiles of the propensity score distribution (Tipton, 2013). Figure 1 shows the distribution of propensity score logits (logit(s(**X**)) = log(s(**X**)/1-s(**X**))) for the Indiana CRT. The experimental sample permitted only three equally sized strata. These strata represent coarse matches between the volunteer and non-volunteer schools based on the covariates used to estimate the propensity scores. As a statistical tool, subclassification shares the same advantages with stratification methods by improving the precision of estimates when schools in the same stratum are more similar in the matched covariates than between strata.

FIGURE 1

Because the bounds are estimated nonparametrically, they can also be computed for subclasses of the data to derive stratum-specific interval estimates. In our final framework, we propose a combined approach where the previous no-assumptions, bounded sample variation with treatment randomization and MTR frameworks are applied with subclassification to derive stratum-specific interval estimates of the PATE. Note that sampling ignorability is not invoked here because the extended application is based on subclassification as a matching method. Given $k$ propensity score strata, the bounds [$PATE_j^L$, $PATE_j^U$] are now estimated using the empirical distribution *($Y_j$, $W_j$, $Z_j$),* for *$j = 1, …, k$*. Since the outcome $Y$ is bounded by the same parameters in each stratum, the interval estimates have the same form as the bounds based on the original sample. Furthermore, the nonparametric method of deriving the bounds offers a flexibility that can easily be extended to any number of subclasses.

**Application to the Indiana CRT**

We now apply the assumptions frameworks to the Indiana CRT. Tables 1 and 2 provide the bounds under the two alternative assumptions and the no-assumptions cases using the experimental and stratified Indiana CRT samples, respectively. The PATE is defined as $P(Y(1)=1) – P(Y(0)=1)$, the difference between the proportion of "Pass" schools if all schools would have implemented the assessment system and the proportion of "Pass" schools if all schools would not have implemented the assessment system. To illustrate this application, fourth-grade scores were used so that the bounds were estimated using $N= 1,029$ fourth-grade-serving schools, a subset of the original 1,514 K–8 schools. From the experimental sample, the conditional probabilities of treatment assignment were estimated as $P(W=1|Z=1) = 34/56 = 0.61$ and $P(W=0|Z=1) = 22/56 = 0.39$. The probability of selection into sample is given by $P(Z=1) = 56/1029 \approx 0.05$ and $P(Z=0) = 1 – P(Z=1)$. The reduced-interval framework requires additional information to estimate $P(W=0, Z=0) = P(W=0|Z=0)*P(Z=0)$. For the purpose of illustrating the bounds and for comparing the assumptions, we let $P(W=0|Z=0) = 0.5$, signifying that schools not in the sample would be randomly assigned to a treatment condition if they participated in the study. Although the probability of receiving either treatment condition was not 0.5 in the actual experimental sample, we use 0.5 for the non-sample schools as a plausible value for this probability in randomized studies with two treatment groups.

TABLE 1

A natural starting place for estimating the PATE is to consider the inferences when no assumptions, beyond SUTVA, strongly ignorable treatment assignment and perfect compliance, are made. The first set of bounds in Table 1 shows the interval estimates under "Treatment Randomization." Here, we estimate the PATE solely using the bounded outcome $Y$ and the randomized nature of the treatment assignment. As shown, the bounds are distinctly uninformative, with the interval nearly spanning the [-1, 1] range for both ELA and Math so that the sign of the PATE cannot be identified. Without any additional information, we cannot determine if the benchmark assessment system had an impact on student achievement. Under the reduced-interval case, incorporating the population data frame narrows the interval estimate. For ELA, the interval shrinks from [-0.93, 0.96] under the full-interval framework to [-0.89, 0.54] under the reduced-interval framework, a 25% shrinkage in width, so that the expected difference in proportion of "Pass" schools is now between -0.89 and 0.54 in the latter framework. Analogous results are seen for Math. Since the probability of sample selection is small, with P(*Z=1*) = 0.05, the intervals include zero.

Some assumptions are thus needed to determine if the benchmark assessment system significantly impacted changes in achievement scores for the study and whether the results generalize to the population of Indiana schools. An important question is, which assumption (if any) is plausible for the Indiana CRT? Sampling ignorability requires that s(**X**) contain all covariates that moderate treatment effects and affect sample selection. In addition, every school in the population must have a comparable school in the CRT example. Although the first part of the assumption is difficult to check, Figure 1 suggests that the second part of the assumption may not be met. In particular, Stratum 3 illustrates that there are some population schools whose sampling propensity scores lie beyond the range of the propensity scores in the sample. This suggests that there exist some schools in the population that may not have comparable schools in the sample and that the plausibility of sampling ignorability is suspect.

We then turn to the first of the alternative assumptions, bounded sample variation, and discuss its plausibility in the Indiana CRT and its tradeoffs. Bounded sample variation assumes that the average difference in expected outcomes between the sample and population is bounded. Determining this assumption's plausibility becomes a question of how well the sample matches the population and assuming that the similarity between sample and population translates into small differences in expected potential outcomes. This similarity has been studied in the literature using balance statistics among covariates (as mentioned in the "Choice of λ" section), the distribution of propensity scores (Stuart et al., 2011), through a generalizability index based on the distribution of propensity scores (Tipton, 2014) and by comparing these measures of similarity to what one would expect in probability samples (Tipton et al., 2016). How similar is the CRT sample to the Indiana population? We refer to the results in Tipton et al. (2016) because measures of similarity between the sample and population were calculated for the same Indiana CRT. For this example, Tipton et al. (2016) compared the balance statistics of the sample with simulated random samples and found that, among the 14 covariates studied, only five had unusually large ASMD when compared to a random sample. Comparisons of the overlap in the propensity score distribution and the generalizability index illustrated that the Indiana CRT sample was not very different from a random sample of similar size. From these assessments, the Indiana CRT sample was considered to be "like" a random sample from the population despite the volunteer nature of sample selection.

Since the CRT sample was considered "like" a probability sample, it potentially shares the advantages of probability sampling. In particular, as a probability sample, we would not expect systematic differences in balance among the covariates, both observable and unobservable. Bounded sample variation is therefore a plausible assumption for the Indiana CRT since, being close to a probability sample, differences between the expected outcomes among the sample and the population would therefore be small on average.

Assuming bounded sample variation, we estimate the bounds for the PATE under the condition that the differences in expected outcomes are small and bounded. We choose two values of $\lambda$ corresponding to the variance of $Y$ and the ASMD of a pre-test covariate, as suggested in the "Choice of $\lambda$" section. These choices give us the bounds in the second rows for each subject in Table 1. For ELA, $\lambda = 0.3$ corresponds to the variance of $Y$ and $\lambda = 0.5$ corresponds to the ASMD of the pretest scores. The values $\lambda = 0.1$ and 0.6 were chosen similarly for Math. For both subjects, the smaller values of $\lambda$ represent a smaller difference in the expected potential outcomes between the volunteer and non-volunteer schools. When $\lambda = 0.3$, the expected difference in proportion of "Pass" schools ranges from -0.31 to 0.83 for ELA, a much tighter interval compared to the treatment randomization case. For Math, the smaller $\lambda$ value of 0.1 actually allows us to identify the sign of the PATE, with the interval estimate ranging from 0.02 to 0.40 so that a positive treatment effect is estimated. This interval suggests that the difference in expected proportions of "Pass" schools between the treatment and control schools ranges from 0.02 to 0.40 so that the benchmark assessment system appears to have a positive impact. These intervals are significantly widened when the value of $\lambda$ increases to 0.5 and 0.6 for ELA and Math, respectively. Since $\lambda$ affects both the lower and upper bounds of $P(Y(1))$ and $P(Y(0))$, the impact on the bounds on the PATE is more pronounced after taking the difference. For the reduced-interval framework, the upper bound PATE$^U$ is much smaller than that of the full-interval framework since the difference $P(Y(1))^L - P(Y(0))^U$ is now taken with a larger $P(Y(0))^U$. With exception to Math with $\lambda = 0.1$, the interval estimates under bounded sample variation still suggest an insignificant PATE.

Bounded sample variation offers a flexibility in estimating the PATE where the difference in expected outcomes between the volunteer and non-volunteer schools is not restricted to be zero. This assumption offers more credibility to inferences if researchers determine that the sample is similar enough to the population or similar enough to a probability sample that the difference in expected outcomes is small. However, like sampling ignorability, bounded sample variation cannot be verified empirically and its application relies on the choice of $\lambda$. While prior data has been used in studies like Manski and Pepper (2015) to empirically choose $\lambda$ to be consistent with the assumption, it is difficult to use prior data in studies like the Indiana CRT in which the study was run for 1 year. Furthermore, determining the plausibility of this assumption with balance statistics is based on the observable covariates and, consequently, has similar concerns with sampling ignorability.

Is MTR a plausible assumption for the Indiana CRT? If MTR holds, the benchmark assessment system improves student outcomes and at worst, has no impact. For the Indiana CRT, we argue that MTR is also plausible but more appropriate than bounded sample variation. Although MTR, too, cannot be empirically validated, its plausibility can be suggested from the logic model for the intervention and even from evidence from pilot studies. In our empirical example, the Indiana State Board of Education planned to use the assessment system to "encourage the advanced and gifted child, drive progress in the student who is ready, and accelerate progress for the student whose learning reflects gaps in preparation and readiness" (Indiana State Board of Education, 2006, pp. 11–12). Like other studies in the use of interim assessments, the system used in the Indiana CRT was designed to identify areas of improvement and use the information to implement instructional strategies to bring about improvement in student outcomes (Konstantopoulos et al., 2013). Since the benchmark assessment system was conceived on the idea that the interim assessments would be instrumental in bringing about improvement in academic outcomes, MTR is a plausible assumption. Because the literature on the use of interim assessments supports the belief that these interventions at least do no harm, we argue that MTR is more appropriate than bounded sample variation in this example. During the pilot year of the CRT, Konstantopoulous et al. (2013) found positive, though insignificant, treatment effects for a majority of the grades among the experimental schools, which lends additional support for MTR.

Under MTR, the expected difference in proportion of "Pass" schools is given in the third rows for ELA and Math in Table 1. Note that all the MTR bounds lie to one side of zero by assumption. The upper bounds using the largest value of $PATE^U$ are largely similar to those under treatment randomization and illustrate that in the absence of information on the sample counterfactuals, the range of values for the PATE is wide. With the MTR bounds using the smallest value of $PATE^U$, the PATE is 0.02 based on the experimental data for both subjects. This upper bound (based on the smallest $PATE^U$) increases to 0.07 and 0.09 for ELA and Math, respectively, when it includes $P(Y(0)=0|W=0, Z=0)$. However, since the difference between 0.02 and 0.07 for ELA is small, $P(Y(0)=0|W=0, Z=0)$ is small, which implies that the proportion of "Not Pass" schools in the population is small. Importantly, because the PATE is a function of $P(Z=1)$, if this proportion is small, the bound $PATE^U$ will also be small. Although the MTR bounds include zero by design, the magnitude of the smallest $PATE^U$ for both subjects suggests that, using the realized outcomes, large values of the PATE can be ruled out even if small insignificant treatment effects cannot be excluded.

MTR would be more appropriate for interventions that are theoretically intended to produce positive impacts. However, there are two tradeoffs. First, unlike bounded sample variation, the bounds under MTR contain zero by assumption so that an insignificant PATE is never ruled out. For this reason, the extent to which the bounds are informative involves comparing the upper bound $PATE^U$. Second, for generalization problems in particular, the bounds are still a function of the unobservable sample counterfactuals so that additional assumptions are needed to tighten the range of values for $PATE^U$.

TABLE 2

**Indiana CRT Bounds by Stratum**

Thus far, the interval estimates under each framework suggest an insignificant treatment effect so that the difference in proportion of "Pass" schools is not significantly different between treatment and control schools. We now consider how these interval estimates compare under the two alternative assumptions when they are estimated in each of the propensity score strata in which the sample and non-sample schools are matched on the observable covariates. Combining the three assumptions framework with subclassification provides a way of observing differences in inferences among schools in individual subclasses. We estimated the propensity scores in this example using a logistic regression model based on covariates from the Indiana population data frame, which included continuous variables such as pre-test measures and binary measures such as Title I status. Table 2 provides the bounds for the PATE under the three frameworks of treatment randomization, bounded sample variation and MTR for ELA and Math. Note that the stratum-specific sample sizes are smaller under this approach, with Stratum 3 containing only two experimental schools, one treatment and one control.

Since subclassification creates matched subgroups of schools under this "fusion" approach, the improved matches strengthen the plausibility of bounded sample variation. Within each subclass, the average difference in observable covariates between the sample and non-sample is minimized. If this translates into small differences in expected outcomes, bounded sample variation may yield bounds that are potentially more informative than those based on the sample as a whole.

From Table 2, the stratum-specific bounds under treatment randomization are largely similar to those based on the entire experimental sample and again are uninformative for both subjects. The reduced-interval bounds under treatment randomization have similar widths as under the original sample, but slight differences can be seen among the strata. For example, the difference in proportion of Pass schools for ELA and Math in Stratum 1 ranges from approximately -0.85 to 0.53 under the reduced-interval framework, but the lower bound decreases to about -0.90 in Stratum 2.

The differences among bounds are seen more distinctly with bounded sample variation and MTR. Using the same bounding parameters as with the original sample, the interval estimates under bounded sample variation suggest larger differences among the strata. In Math, for example, the sign of the PATE is identified in Stratum 1 when $\lambda = 0.1$ with an interval estimate of [0.12, 0.49]. This implies that the expected difference in proportion of "Pass" schools among the treatment and control schools is between 0.12 and 0.49 for schools in this stratum. However, the bounds in Stratum 2 imply an insignificant PATE, which illustrate potential differences in inferences among the strata.

TABLE 3

As a comparison, Table 3 gives the point estimates of the PATE under "no weighting," inverse propensity weighting (IPW), and subclassification for ELA and Math. These point estimates are given assuming all of the conditions of sampling ignorability hold. The "no weighting" case refers to the estimate of the SATE, which is likely biased as an estimate for the PATE in the absence of probability sampling. The point estimates for ELA and Math are all insignificant, a result that is largely consistent with the bounds provided. While the interval estimates for Math show a positive PATE under $\lambda = 0.1$ with bounded sample variation, the lower bound of 0.02 suggests that small insignificant treatment effects are possible. With exception to IPW, the point estimates under "no weighting" and subclassification are similar in magnitude to the smallest PATE$^U$ under MTR, further supporting insignificant results.

In the original analysis, Konstantopoulos et al. (2013) found significant treatment effects for fourth-grade ELA using a two-level hierarchical linear model with covariates. Using standardized continuous ISTEP+ scores, a significant PATE of 0.135 (0.057 standard error) was found based on the experimental sample with a model that included school- and student-level covariates. Importantly, the PATE using the model with treatment alone was not significant (PATE = 0.087 with 0.111 standard error). Although this point estimate is not directly comparable to the bounds based on binary outcomes provided here, it is important to note that the significance of the estimate and the resulting inferences depended on the choice of model.

Several items should be noted following these comparisons. First, assumptions play a significant role in determining the extent to which estimates are informative in generalization studies with non-probability samples. Bounded sample variation alone tightened the interval estimate of [-0.93, 0.96] to [-0.31, 0.83] under the full interval framework for ELA. Second, differences in assumptions are reflected in the various magnitudes of the interval estimates, which can lead to different inferences. If treatment is assumed to have a positive impact (under MTR), the smallest PATE$^U$ is clearly different from the lower bounds under bounded sample variation where the differences in expected outcomes are allowed to be positive or negative. Lastly, different methods using the same assumptions may lead to different estimates of the parameter. Table 3 illustrates that the three methods of estimating the PATE give slightly different values even though sampling ignorability assumptions are made. Additionally, although an insignificant PATE is the consistent result among the interval estimates and point estimates in our example, it is important to note that this may not be true in other studies. From our comparisons, we highlight the need for thoughtful consideration of assumptions, their plausibility, and their implications for inference.

**Discussion & Conclusion**

This paper compares sampling ignorability with two alternative assumptions in the context of causal generalization from non-probability samples. Sampling ignorability is one of several key assumptions needed to point identify the PATE, but as our empirical example suggests, situations in which it is violated may occur in practice. Our comparison of sampling ignorability with bounded sample variation and MTR illustrates that data alone are not sufficiently informative of the PATE and that there are tradeoffs to invoking different assumptions. For example, bounded sample variation may be plausible based on measures of balance between the sample and population, but the assumption involves choosing a value for $\lambda$, which can be challenging.

A common difficulty among each of the assumptions (sampling ignorability, bounded sample variation, and MTR) is that they are all are untestable by the data. With MTR, the data and the logic model for the intervention can at most suggest its plausibility but cannot provide validation. The researcher is then left with deciding which assumptions are plausible based on which seem most credible and consistent with the data at hand and with the theoretically proposed impact. Our focus on MTR and its plausibility in the Indiana CRT, for example, was motivated by the Indiana State Board of Education's proposal to use interim assessments to improve academic outcomes. Prior information and theoretical evidence of the impact of an intervention are important in deciding the plausibility of assumptions, a task that would otherwise be difficult in practice.

Generalizations with non-probability samples inevitably involve a discussion of the extent to which a self-selected sample differs from the population. The role that all assumptions in generalization problems plays is to make conjectures about this difference. The literature on the performance of estimators when assumptions do not hold is extensive and has been explored in simulation (see Kern, Stuart, Hill, & Green, 2016). However, it is much more challenging to design studies to assess whether sampling ignorability or bounded sample variation hold in practice. The comparison presented in this paper combines two perspectives of inference by assessing the differences between point identifying and partially identifying assumptions. While interval estimates do not substitute for the point estimates used to inform policy, we present their application as a way to highlight the importance of considering the plausibility and credibility of different assumptions, particularly when they lead to potentially different inferences. The assumptions explored in this article are only a subset of the alternatives to sampling ignorability. Future work should continue to explore other alternative assumptions, some of which may impose different constraints than sampling ignorability, and assess their plausibility and their effect on inferences. Additionally, assumptions that specify different distributional conditions on the treatment effects should also be explored, possibly through simulation.

While not addressed here, it is important to note that the interval estimates provided in this article do not incorporate standard errors. Previous research has explored asymptotically valid confidence intervals (Imbens & Manski, 2004) and estimation methods for intersection bounds (interval estimates) that provide asymptotically valid inferences (Chernozhukov, Lee, & Rosen, 2013). However, the focus of these studies was on the theoretical development of large sample inferential methods and not necessarily on sampling error. Future research should explore methods of incorporating standard errors for partially identified parameters, particularly when interval estimates are used for inference. This incorporation will be important in fields such as meta-analysis where knowledge of sampling errors affect the synthesis of results.

Figure 1. Distribution of Propensity Score Logits for Indiana CRT

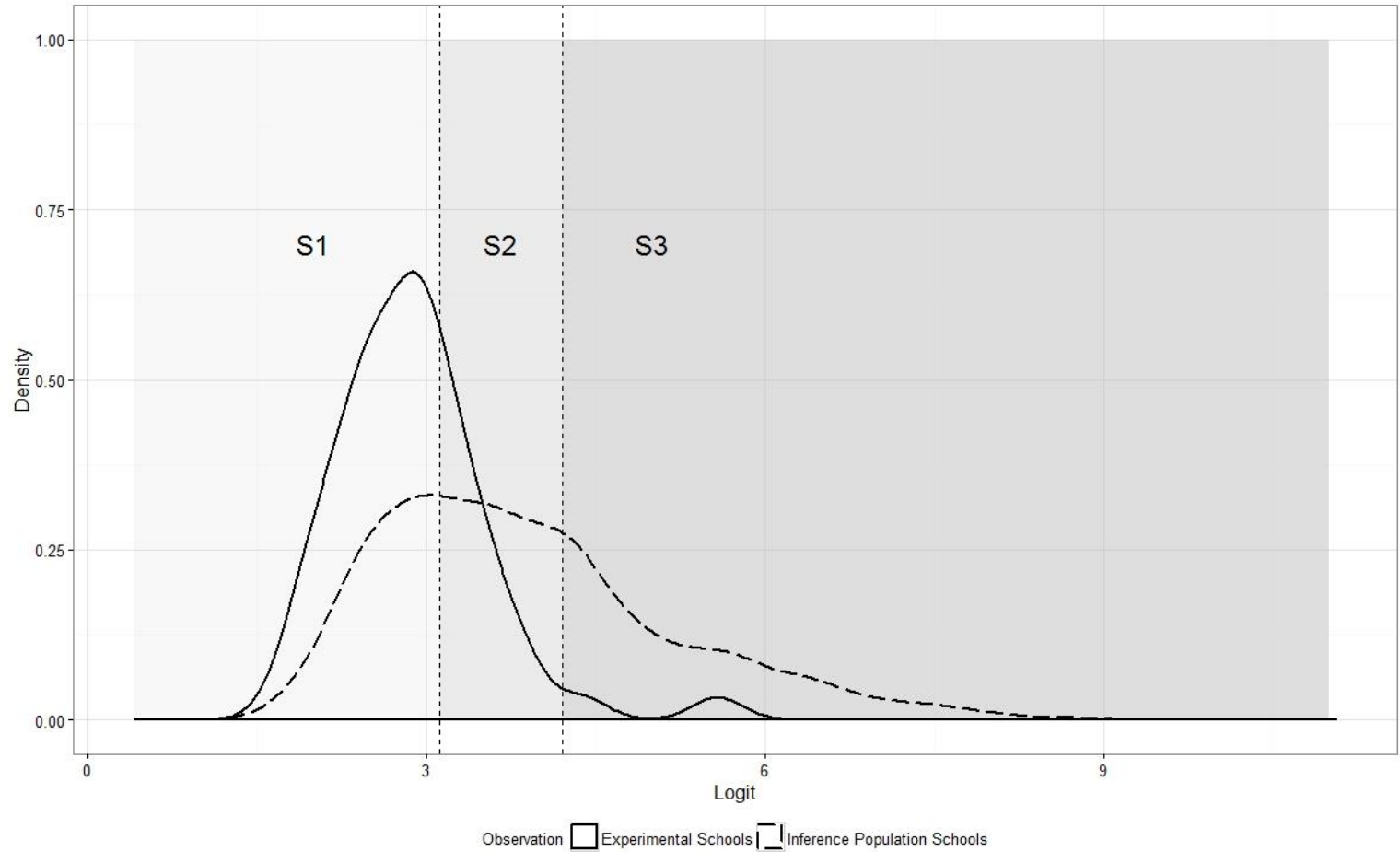

Note: S1: Stratum 1; S2: Stratum 2; S3: Stratum 3

Table 1. Bounds on PATE for Indiana CRT

| | Identifying Assumptions | Full-Interval | | Reduced-Interval* | |
|---|---|---|---|---|---|
| ELA | Treatment Randomization | [-0.93, 0.96] | | [-0.89, 0.54] | |
| | Bounded Sample Variation | $\lambda = 0.3$ [-0.31, 0.83] | $\lambda = 0.5$ [-0.69, 0.99] | $\lambda = 0.3$ [-0.43, 0.43] | $\lambda = 0.5$ [-0.71, 0.71] |
| | Monotone Treatment Response | Sample MTR [0.00, 0.02] [†] [0.00, 0.97] [††] | | Population MTR [0.00, 0.07] [†] [0.00, 0.54] [††] | |
| Math | Treatment Randomization | [-0.93, 0.96] | | [-0.87, 0.55] | |
| | Bounded Sample Variation | $\lambda = 0.1$ [0.02, 0.40] | $\lambda = 0.6$ [-0.93, 0.99] | $\lambda = 0.1$ [-0.18, 0.11] | $\lambda = 0.6$ [-0.89, 0.82] |
| | Monotone Treatment Response | Sample MTR [0.00, 0.02] [†] [0.00, 0.97] [††] | | Population MTR [0.00, 0.09] [†] [0.00, 0.56] [††] | |

\* Reduced Interval bounds derived using P(*W*=0|*Z*=0) = 0.5.
[†] Smallest value of $PATE^U$ is used.
[††] Largest value of $PATE^U$ is used.

Table 2. Bounds on PATE for Indiana CRT by Stratum

| | | Randomized Treatment | | Bounded Sample Variation | | | | Monotone Treatment Response | |
|---|---|---|---|---|---|---|---|---|---|
| | | Full Interval | Reduced Interval* | Full-Interval | | Reduced-Interval* | | Sample MTR | Population MTR |
| | | | | $\lambda=\lambda_1$‡ | $\lambda=\lambda_2$‡ | $\lambda=\lambda_1$‡ | $\lambda=\lambda_2$‡ | | |
| ELA | S1 | [-0.90, 0.95] | [-0.86, 0.52] | [-0.22, 0.89] | [-0.59, 0.99] | [-0.36, 0.48] | [-0.63, 0.75] | [0.00, 0.03] [0.00, 0.96] | [0.00, 0.07] [0.00, 0.53] |
| | S2 | [-0.94, 0.95] | [-0.92, 0.50] | [-0.50, 0.63] | [-0.88, 0.99] | [-0.61, 0.24] | [-0.89, 0.53] | [0.00, 0.02] [0.00, 0.96] | [0.00, 0.04] [0.00, 0.51] |
| | S3 | [-0.99, 0.99] | [-0.89, 0.60] | [-0.10, 0.99] | [-0.49, 0.99] | [-0.34, 0.55] | [-0.64, 0.85] | [0.00, 0.004] [0.00, 0.996] | [0.00, 0.11] [0.00, 0.60] |
| Math | S1 | [-0.90, 0.95] | [-0.85, 0.53] | [0.12, 0.49] | [-0.80, 0.99] | [-0.11, 0.17] | [-0.80, 0.86] | [0.00, 0.03] [0.00, 0.96] | [0.00, 0.08] [0.00, 0.54] |
| | S2 | [-0.95, 0.94] | [-0.90, 0.51] | [-0.25, 0.13] | [-0.99, 0.99] | [-0.40, 0.11] | [-0.99, 0.60] | [0.00, 0.01] [0.00, 0.96] | [0.00, 0.06] [0.00, 0.53] |
| | S3 | [-0.99, 0.99] | [-0.85, 0.64] | [0.30, 0.70] | [-0.59, 0.99] | [-0.006, 0.29] | [-0.75, 0.99] | [0.00, 0.004] [0.00, 0.996] | [0.00, 0.14] [0.00, 0.64] |

*Reduced Interval bounds derived using $P(W=0|Z=0) = 0.5$.
‡ For ELA, $\lambda_1 = 0.3$ and $\lambda_2 = 0.5$. For Math, $\lambda_1 = 0.1$ and $\lambda_2 = 0.6$.
The top MTR bounds are based on the smallest value of PATE$^U$ while the bottom MTR bounds are based on the largest value of PATE$^U$.

Table 3. Point Estimates of the PATE for Indiana CRT

|  | Method | Estimate | Standard Error |
|---|---|---|---|
| ELA | No Weighting | 0.048 | 0.038 |
|  | IPW | 0.128 | 0.116 |
|  | Subclassification | 0.056 | 0.056 |
| Math | No Weighting | 0.095 | 0.051 |
|  | IPW | 0.158 | 0.117 |
|  | Subclassification | 0.097 | 0.063 |

## Online Supplementary Materials

This section contains the proofs of sharpness of the bounds and sample R code for computing the interval estimates.

## Proofs of Sharpness of Bounds

### Treatment Randomization

We follow the structure of the proof in Heckman and Vytlacil (2001) in proving the sharpness of the bounds in (7) under treatment randomization. Following the logic of Heckman and Vytlacil (2001), we first show that for any point $b$ in the bounds [PATE$^L$, PATE$^U$] in (7), there exists a distribution that (i) is consistent with the observed data; (ii) is consistent with the assumptions stated in the *Treatment Randomization* section and; (iii) such that the PATE $E(Y(1) - Y(0))$ evaluated under the distribution equals $b$. The proof proceeds by constructing a distribution that satisfies each of the three conditions for any given $b$ in [PATE$^L$, PATE$^U$].

*Proof:*

Let P(Z) denote the probability distribution of the sample selection indicator Z and let $P_z$ denote the support of P(Z). Let $p_z^u = \sup P_z$ and $p_z^l = \inf P_z$. Let $w$ denote a realization of the treatment assignment variable W and let $z$ denote a realization of the sample selection variable Z. For any random variable X, let $F^0_X$ be the "true" cumulative distribution function (CDF) of X and let $F^0_{X|Y}(.\,|y)$ denote the true CDF of X conditional on $Y = y$. We assume that $Y(1), Y(0)$ are bounded by the same bounds so that for almost every $w \in W$ and $z \in Z$, there exists $Y^L$, $Y^U$ such that $P(Y^L \leq Y(w) \leq Y^U | W = w, Z = z) = 1$. Let $y(0)$ and $y(1)$ denote a realization of the respective potential outcomes. As described above, let $b$ denote any given element in the bounds [PATE$^L$, PATE$^U$] from (7).

Note that we can rewrite any element $b$ as the following:

$$b = p_z^u \big( E(Y(1)|P(Z) = p_z^u, W = 1) \big) + (1 - p_z^l) q^1$$

$$- (1 - p_z^l) \big( E(Y(0)|P(Z) = p_z^l, W = 0) \big) - p_z^l q^0$$

for some $q^0, q^1$ such that $Y^L \leq q^w \leq Y^U$ for $w \in W$.

For $z$ in the support of Z, define

$$F_{Y(1)|Z}(y(1)|z) = \begin{cases} F^0{}_{Y(1)|Z}(y(1)|z) & \text{if } z \leq p_z{}^u \\ \mathbf{1}\{Y(1) \geq q^1\} & \text{if } z > p_z{}^u \end{cases}$$

$$F_{Y(0)|Z}(y(0)|z) = \begin{cases} F^0{}_{Y(0)|Z}(y(0)|z) & \text{if } z \geq p_z{}^l \\ \mathbf{1}\{Y(0) \geq q^0\} & \text{if } z < p_z{}^l \end{cases}$$

where $\mathbf{1}\{.\}$ is an indicator function.

Define

$$F_{Y(1),Y(0),W,Z}(y(1),y(0),w,z)$$

$$= \int \int_0^z F_{Y(1)|Z}(y(1)|u_z) * F_{Y(0)|Z}(y(0)|u_z) dF_Z(u_z) \mathbf{1}\{u_w \leq w\} dF^0{}_W(u_w)$$

Here, $F^0{}_Z$ and $F^0{}_W$ are the "true" CDFs of $Z$ and $W$. Under the given construction, $F$ satisfies the following conditions:

1) $F$ is a proper CDF.

2) The potential outcomes $Y(1)$, $Y(0)$ are bounded by $Y^L$, $Y^U$.

3) The treatment assignment variable $W$ is independent of the potential outcomes $Y(1)$, $Y(0)$.

By construction, $F_Z(z) = F^0{}_Z(z)$ and using $F_{Y(1)|Z}(y(1)|z) = F^0{}_{Y(1)|Z}(y(1)|z)$ for $z \leq p_z{}^u$, we have:

$$F_{Y(1)|Z,W}(y(1)|z, w = 1) = \frac{1}{P(z)} \int_0^{P(z)} F^0{}_{Y(1)|Z}(y(1)|z) dF^0{}_Z(z) = F^0{}_{Y(1)|Z,W}(y(1)|z, w = 1)$$

for $z$ in the support of $(Z/W=1)$. An analogous argument gives:

$$F_{Y(0)|Z,W}(y(0)|z, w = 0) = \frac{1}{P(z)} \int_0^{P(z)} F^0{}_{Y(0)|Z}(y(0)|z) dF^0{}_Z(z) = F^0{}_{Y(0)|Z,W}(y(0)|z, w = 0)$$

for $z$ in the support of $(Z|W=0)$. Given these two results, we have:

$$F_{Y,Z,W}(y,z,w) = F^0{}_{Y,Z,W}(y,z,w)$$

where $Y = Z*(WY(1)+(1-W)Y(0)) + (1-Z)*(WY(1)+(1-W)Y(0))$ so that $F$ is observationally equivalent to $F^0$. The PATE, which is the expected value of $Y(1) - Y(0)$, under $F$ is given by:

$$E(Y(1) - Y(0)|Z)$$

$$= \int \left( \int y(1) dF_{Y(1)|Z}(y(1)|z) \right) dF^0{}_Z(z)$$

$$- \int \left( \int y(0) dF_{Y(0)|Z}(y(0)|z) \right) dF^0{}_Z(z)$$

$$= P(z \leq p_z{}^u) \int \left( \int_0^{p_z{}^u} y(1) dF^0{}_{Y(1)|Z}(y(1)|z) \right) dF^0{}_Z(z) + P(z > p_z{}^u) q^1$$

$$- P(z > p_z{}^l) \int \left( \int_{p_z{}^l}^1 y(0) dF^0{}_{Y(0)|Z}(y(0)|z) \right) dF^0{}_Z(z) - P(z \leq p_z{}^l) q^0$$

$$= p_z{}^u E(Y(1)|P(Z) = p_z{}^u, W = 1) + (1 - p_z{}^u) q^1 - p_z{}^l E(Y(0)|P(Z) = p_z{}^l, W = 0)$$
$$- p_z{}^l q^0$$

$$= b$$

The expected value of the PATE, $Y(1) - Y(0)$ under $F$ equals $b$, and since $F$ satisfies all of the conditions of the given example and is observationally equal to the true CDF $F^0$, the point $b$ must be contained in any bounds on the average treatment effect. Since this is true for any point $b \in$ [PATE$^L$, PATE$^U$] we have that every point in [PATE$^L$, PATE$^U$] must be contained in any bounds on the average treatment effect so that the bounds [PATE$^L$, PATE$^U$] are tight under the given information structure.

The bounds above are given for an outcome $Y$ that is bounded between $Y^L$ and $Y^U$. In the context of the Indiana CRT, $Y^L = 0$ and $Y^U = 1$ and making these substitutions for $b$ above gives us the bounds in (7) under the full-interval framework. Substituting $Y^L$ and $Y^U$ for all of the unobservable sample counterfactuals except for $E(Y(0)/W=0, Z=0)$ under the reduced-interval framework yields analogous bounds for the PATE in (7).

*Bounded Sample Variation and Treatment Randomization*

We again assume that the potential outcomes $Y(1)$, $Y(0)$ are bounded by the same lower and upper bound, $Y^L$, $Y^U$. For $w \in W$ and $z \in Z$, bounded sample variation stipulates the following conditions:

$E(Y(w)/W=w, Z=1) - \lambda \leq E(Y(w)/W=w, Z=0) \leq E(Y(w)/W=w, Z=1) + \lambda$ and

$E(Y(w)|W \neq w, Z=1) - \lambda \leq E(Y(w)|W \neq w, Z=0) \leq E(Y(w)|W \neq w, Z=1) + \lambda$

Under the given information structure, the bounds under this assumption and treatment randomization are sharp and improve upon the bounds under treatment randomization if

$(Y^U - Y^L) > E(Y(1)/W=1, Z=1) - E(Y(0)/W=0, Z=1) + 2\lambda)$ and

$(Y^L - Y^U) < E(Y(1)/W=1, Z=1) - E(Y(0)/W=0, Z=1) - 2\lambda)$

*Proof:*

Using the law of iterated expectations, the expected values of the potential outcomes, $E(Y(1))$, $E(Y(0))$, can be written as:

$E(Y(1)) = E(Y(1)/W=1, Z=1)* P(W=1, Z=1) + E(Y(1)/W=0, Z=1)*P(W=0, Z=1) +$

$\quad E(Y(1)/W=1, Z=0)*P(W=1, Z=0) + E(Y(1)/W=0, Z=0)*P(W=0, Z=0)$

$E(Y(0)) = E(Y(0)/W=1, Z=1)* P(W=1, Z=1) + E(Y(0)/W=0, Z=1)*P(W=0, Z=1) +$

$\quad E(Y(0)/W=1, Z=0)*P(W=1, Z=0) + E(Y(0)/W=0, Z=0)*P(W=0, Z=0)$

Under treatment randomization, the expected values of the potential outcomes are given by:

$E(Y(1)) = E(Y(1)/W=1, Z=1)*P(Z=1) + E(Y(1)/W=1, Z=0)* P(W=1, Z=0) +$

$\quad E(Y(1)/W=0, Z=0)*P(W=0, Z=0)$

$E(Y(0)) = E(Y(0)/W=0, Z=1)*P(Z=1) + E(Y(0)/W=1, Z=0)* P(W=1, Z=0) +$

$\quad E(Y(0)/W=0, Z=0)*P(W=0, Z=0)$

Under bounded sample variation,

$E(Y(1)/W=1, Z=1) - \lambda \leq E(Y(1)/W=1, Z=0) \leq E(Y(1)/W=1, Z=1) + \lambda$ and

$E(Y(1)/W=1, Z=1) - \lambda \leq E(Y(1)/W=0, Z=0) \leq E(Y(1)/W=1, Z=1) + \lambda$ where

$E(Y(1)/W=0, Z=1) = E(Y(1)/W=1, Z=1)$

under treatment randomization. Similarly,

$E(Y(0)/W=0, Z=1) - \lambda \leq E(Y(0)/W=0, Z=0) \leq E(Y(0)/W=0, Z=1) + \lambda$ and

$E(Y(0)/W=0, Z=1) - \lambda \leq E(Y(0)/W=1, Z=0) \leq E(Y(0)/W=0, Z=1) + \lambda$

This implies that,

$E(Y(1)/W=1, Z=1)*P(Z=1) + (E(Y(1)/W=1, Z=1) - \lambda)*(1- P(Z=1))$

$\leq E(Y(1)) \leq$

$E(Y(1)/W=1, Z=1)*P(Z=1) + (E(Y(1)/W=1, Z=1) + \lambda)*(1- P(Z=1))$ and

E(*Y(0)*/*W=0, Z=1*)*P(*Z=1*) + (E(*Y(0)*/*W=0, Z=1*) – $\lambda$)*(1- P(*Z=1*))

$\leq$ E(*Y(0)*) $\leq$

E(*Y(0)*/*W=0, Z=1*)*P(*Z=1*) + (E(*Y(0)*/*W=0, Z=1*) + $\lambda$)*(1- P(*Z=1*))

The PATE is therefore bounded by

PATE$^L$ $\leq$ E(*Y(1) – Y(0)*) $\leq$ PATE$^U$

where

PATE$^L$ = (E(*Y(1)*/*W=1, Z= 1*) – E(*Y(0)*/*W=0, Z=1*))*P(*Z=1*) + (E(*Y(1)*/*W=1, Z=1*) –

E(*Y(0)*/*W=0, Z=1*) - 2 $\lambda$)*(1-P(*Z=1*))

PATE$^U$ = (E(*Y(1)*/*W=1, Z= 1*) – E(*Y(0)*/*W=0, Z=1*))*P(*Z=1*) + (E(*Y(1)*/*W=1, Z=1*) –

E(*Y(0)*/*W=0, Z=1*) + 2 $\lambda$)*(1-P(*Z=1*))

For every value in the interval [PATE$^L$, PATE$^U$], we can construct a distribution of (*Y(1), Y(0), W, Z*) that is consistent with the observed distribution of (*Y, W, Z*) where *Y = Z\*(W\*Y(1)+(1-W)\*Y(0)) + (1-Z)\*(W\*(Y(1)+(1-W)\*Y(0))* and such that the PATE equals the specified value. The bounds [PATE$^L$, PATE$^U$] improve upon the bounds under treatment randomization if

(1) ($Y^L - Y^U$) < (E(*Y(1)*/*W=1, Z=1*) – E(*Y(0)*/*W=0, Z=1*) - 2 $\lambda$)      and

(2) ($Y^U - Y^L$) > (E(*Y(1)*/*W=1, Z=1*) – E(*Y(0)*/*W=0, Z=1*) + 2 $\lambda$)

Given $\lambda$, if (1) and (2) hold, every point in the interval [PATE$^L$, PATE$^U$] must be contained in any bounds on the PATE so that these bounds are tight under the given information structure.

The bounds under the reduced interval framework are derived similarly, but no substitutions are made for E(*Y(0)*/*W=0, Z=0*).

### *Monotone Treatment Response*

Let *W* denote an ordered set of treatments and we assume that *Y(1), Y(0)* are bounded by $Y^L$, $Y^U$. For each school *i* in the population, *i = 1, …, N,* let $w_i$ denote the treatment assigned to school *i*, *Y($w_i$)* is the realized outcome of school *i* under treatment $w_i$ and let $Y^L(w_i)$ and $Y^U(w_i)$ denote the smallest and largest feasible values of *Y($w_i$)*.

*Proof:*

We extend the proof of Manski (2009) for the case of a binary treatment assignment variable *W* and a binary sample selection variable *Z*. We first prove that $Y^L(w_i) \leq Y(w_i) \leq Y^U(w_i)$ are sharp bounds for *Y($w_i$)* for $w_i$ = 0, 1, for all *i* in the population. We then use this result to show that the bounds under MTR in (12) are sharp under the assumption of MTR for every school in the population. MTR stipulates that for all schools $i$, $i=1, ..., N$, given two treatments $w=1$, $w'=0$, with $w, w' \in W$, the following condition holds:

$$(w=1) \geq (w'=0) \rightarrow Y(w=1) \geq Y(w'=0)$$

We define

$$Y^L(w_i) \equiv \begin{cases} Y(w_i) & \text{if } w_i = 0 \\ Y^L & \text{otherwise} \end{cases}$$

and

$$Y^U(w_i) \equiv \begin{cases} Y(w_i) & \text{if } w_i = 1 \\ Y^U & \text{otherwise} \end{cases}$$

Then, by MTR,

$w_i = 0 \rightarrow Y^L \leq Y^L(w_i) \leq Y(w_i) \leq Y(w_i=1)$

$w_i = 1 \rightarrow Y(w_i=0) \leq Y(w_i) \leq Y^U(w_i) \leq Y^U$

This implies that $Y^L(w_i) \leq Y(w_i) \leq Y^U(w_i)$ and since there is no crossover of treatment among schools in the sample and population, the bounds $[Y^L(w_i), Y^U(w_i)]$ are sharp for all schools $i$ in the population. As a result, the bounds $E(Y^L(w)) \leq E(Y(w)) \leq E(Y^U(w))$ are sharp for all $i$ and the empirical evidence and the MTR assumption are consistent with the hypothesis $\{Y(w_i) = Y^L(w_i)\}$ for all $i$ and with the hypothesis $\{Y(w_i) = Y^U(w_i)\}$ for all $i$.

Here, we illustrate that under MTR, the bounds for the PATE given by $0 \leq E(Y(1) - Y(0)) \leq E(Y^U(1)) - E(Y^L(0))$ are sharp.

*Proof:*

Given treatments $w=1$, $w'=0$, MTR stipulates that if $(w=1) \geq (w'=0)$, $Y(w=1) \geq Y(w'=0)$ so that zero is a lower a bound on $E(Y(1)) - E(Y(0))$. The result $E(Y^L(w)) \leq E(Y(w)) \leq E(Y^U(w))$ from the previous section implies that $E(Y^U(1)) - E(Y^L(0))$ is an upper bound.

For school $i$, $i=1, ...N$, MTR of $Y(w)$ implies:

$w'=0 \leq w=1 = w_i \rightarrow Y^L \leq Y^L(w_i=0) \leq Y(w_i=0) \leq Y(w_i=1)$

$w_i = w'=0 \leq w=1 \rightarrow Y(w_i=0) \leq Y(w_i=1) \leq Y^U(w_i=1) \leq Y^U$

Since there are no treatment crossovers in the sample and population, the data and MTR assumption are consistent with the hypothesis $\{Y(w_i = 0) = Y(w_i=1),$ for all $i\}$ and with the hypothesis $\{Y(w_i =1) = Y^U(w_i = 1),\ Y(w_i =0) = Y^L(w_i = 0),$ for all $i\}$ so that the bounds $[0, E(Y^U(1)) - E(Y^L(0))]$ are sharp under the given information structure. Because $Y$ is a function of both $W$ and $Z$, the upper bound $E(Y^U(1)) - E(Y^L(0))$ is also a function of both $W$ and $Z$. This upper bound is decomposed into the conditional probabilities based on $Z = 0, 1$ in the derivation of (11). Since $Y^L = 0$ in our context, the upper bound is given solely by $E(Y^U(1))$.

## Sample R Code to Compute Bounds

*Bounds under Treatment Randomization*

```
# Full Interval Case #

# The following function, "bounds_randtrt," computes the bounds under treatment randomization.

bounds_randtrt<-function(x,y){ # x denotes the sample data, y denotes the population data; out denotes the outcome of interest

  y1w1z1<-sum(x$out[x$trt==1])/nrow(x) # estimates P(Y(1)|W=1,Z=1)

  y0w0z1<-sum(x$out[x$trt==0])/(nrow(x)) # estimates P(Y(0)|W=0,Z=1)

  y0w0z0<-sum(y$out)/((nrow(subset(y,is.na(y$out)==F))))

            # estimates P(Y(0)|W=0,Z=0)

  prob_z1<-nrow(x)/(nrow(subset(y,is.na(y$out)==F))) # estimates P(Z=1)

  prob_z0<-(nrow(subset(y,is.na(y$out)==F))/nrow(subset(y,is.na(y$out)==F)))-
      nrow(x)/(nrow(subset(y,is.na(y$out)==F)))        # estimates P(Z=0)

  LB<-(y1w1z1*prob_z1)-((y0w0z1)*prob_z1 + prob_z0)   # lower bound

  UB<-(y1w1z1)*prob_z1 +prob_z0 -((y0w0zz)*prob_z1)  # upper bound

  bds<-as.matrix(c(LB,UB))

  rownames(bds)<-c("LB","UB")

  return(bds)

}

# Reduced Interval Case #

# The function, "bounds_randtrt_red," computes the bounds under treatment randomization for the reduced interval case.

bounds_randtrt_red<-function(x,y){ # x denotes the sample data, y denotes the population data

  y1w1z1<-sum(x$out[x$trt==1])/nrow(x) # estimates P(Y(1)|W=1,Z=1)

            # estimates P(Y(1)|W=1,Z=1)

  y0w0z1<-sum(x$out[x$trt==0])/(nrow(x)) # estimates P(Y(0)|W=0,Z=1)

            # estimates P(Y(0)|W=0,Z=1)

  y0w0z0<-sum(y$out)/((nrow(subset(y,is.na(y$out)==F))))
```

```
                # estimates P(Y(0)|W=0,Z=0)

  prob_z1<-nrow(x)/(nrow(subset(y,is.na(y$out)==F))) # estimates P(Z=1)

  prob_z0<-(nrow(subset(y,is.na(y$out)==F))/nrow(subset(y,is.na(y$out)==F)))-
           nrow(x)/(nrow(subset(y,is.na(y$out)==F))) # estimates P(Z=0)

  p_w0z0 <- 0.5 # estimates P(W=0|Z=0)

  LB<-(y1w1z1*prob_z1)-((y0w0z1)*prob_z1 + (y0w0z0*p_w0z0*prob_z0) +
      (1-p_w0z0*prob_z0-prob_z1))  # lower bound

  UB<-(y1w1z1)*prob_z1 + prob_z0-((y0w1z0)*prob_z1+y0w0z0*p_w0z0*prob_z0)
           # upper bound

  bds<-as.matrix(c(LB,UB))

  rownames(bds)<-c("LB","UB")

  return(bds)

}
```

*Bounded Sample Variation and Treatment Randomization*

```
lambda<-0.3      # bounding parameter
```

# The following function, "bded_var," computes the bounds under bounded sample variation under treatment randomization with bounding parameter lambda.

```
bded_var<-function(x,y,lambda){

  y1w1z1<-sum(x$out[x$trt==1])/nrow(x)    # estimates P(Y(1)|W=1,Z=1)

  y0w0z1<-sum(x$out[x$trt==0])/(nrow(x))  # estimates P(Y(0)|W=0,Z=1)

  y0w0z0<-sum(y$out)/((nrow(subset(y,is.na(y$out)==F))))

                # estimates P(Y(0)|W=0,Z=0)

  prob_z1<-nrow(x)/(nrow(subset(y,is.na(y$out)==F))) # estimates P(Z=1)

  prob_z0<-(nrow(subset(y,is.na(y$out)==F))/nrow(subset(y,is.na(y$out)==F)))-
           nrow(x)/(nrow(subset(y,is.na(y$out)==F))) # estimates P(Z=0)

  p_w0z0 <- 0.5   # estimates P(W=0|Z=0)

  y1_lower<-y1w1z1*prob_z1 + ((y1w1z1-lambda))*(1-prob_z1)
            # estimates the lower bound of Y(1)
```

```
    y1_upper<-y1w1z1*prob_z1 + ((y1w1z1+lambda))*(1-prob_z1)
            # estimates the upper bound of Y(1)

    y0_lower<-y0w0z1*prob_z1+((y0w0z1-lambda))*(1-prob_z1)
            # estimates the lower bound of Y(0)

    y0_upper<-y0w0z1*prob_z1+((y0w0z1+lambda))*(1-prob_z1)
            # estimates the upper bound of Y(0)

    y0_lower_red<-y0w0z1*prob_z1+y0w0z0*p_w0z0*prob_z0+((y0w0z1-
        lambda))*p_w0z0*(1-prob_z1)

# estimates the lower bound of Y(0) under the reduced interval framework

    y0_upper_red<-y0w0z1*prob_z1+y0w0z0*p_w0z0*prob_z0
        +((y0w0z1+lambda))*p_w0z0*(1-prob_z1)

# estimates the upper bound of Y(0) under the reduced interval framework

    LB<-y1_lower-y0_upper

    UB<-y1_upper-y0_lower        # bounds under the full interval framework

    LB_red<-y1_lower-y0_upper_red

    UB_red<-y1_upper-y0_lower_red     # bounds under the reduced interval
                                        framework

    bds_bdedvar<-as.matrix(c(LB,UB,LB_red,UB_red))

    rownames(bds_bdedvar)<-c("LB","UB","LB_red","UB_red")

    return(bds_bdedvar)

}
```

*Monotone Treatment Response*

```
# The following function, "bounds_mono," computes the bounds under monotone
treatment response.

bounds_mono<-function(x,y){

  y0w1z0<-sum(x$out[x$trt==0]==0)/nrow(x)       # estimates P(Y(0)=0|W=0,Z=0)

  y0w0z0<-sum(y$out)/((nrow(subset(y,is.na(y$out)==F))))
```

```r
                    # estimates P(Y(0)|W=0,Z=0)
    y1w1z1<-sum(x$out[x$trt==1]==1)/nrow(x)        # estimates P(Y(1)|W=1,Z=1)
    

    p_w1z1<-(nrow(x[x$trt==1,]))/nrow(x)           # estimates P(W=1,Z=1)
    p_w0z1<-(nrow(x[x$trt==0,]))/nrow(x)           # estimates P(W=0,Z=1)
    
    
    prob_z1<-nrow(x)/(nrow(subset(y,is.na(y$out)==F))) # estimates P(Z=1)
    prob_z0<-(nrow(subset(y,is.na(y$out)==F))/nrow(subset(y,is.na(y$out)==F)))-
        nrow(x)/(nrow(subset(y,is.na(y$out)==F)))    # estimates P(Z=0)
    
    LB_mono<-0     # lower bound, assumed to be 0
    UB_samp_LB<-y0w0z1*p_w0z1*prob_z1 + y1w1z1*p_w1z1*prob_z1
                   # Sample MTR upper bound using the smallest PATE_U
    UB_samp_UB<-y0w0z1*p_w0z1*prob_z1 + y1w1z1*p_w1z1*prob_z1+prob_z0
                   # Sample MTR upper bound using the largest PATE_U
    UB_pop_LB<-y0w0z1*p_w0z1*prob_z1 + y1w1z1*p_w1z1*prob_z1 +
            y0w0z0*p_w0z0*prob_z0
                   # Population MTR upper bound using the smallest PATE_U
    UB_pop_UB<-y0w0z1*p_w0z1*prob_z1 + y1w1z1*p_w1z1*prob_z1 +
         y0w0z0*p_w0z0*prob_z0+(1-p_w0z0*prob_z0-p_w0z1*prob_z1-p_w1z1*prob_z1)
                   # Population MTR upper bound using the smallest PATE_U
    bds_mono<-as.matrix(c(UB_samp_LB,UB_samp_UB,UB_pop_LB,UB_pop_UB))
    rownames(bds_mono)<-
c("SATE_MTR_LB","SATE_MTR_UB","PATE_MTR_LB","PATE_MTR_UB")
    return(bds_mono)
}
```